# Interfacial Effects on the Optical Properties of CdTe/CdS Quantum Dots

Girija Gaur[1], Dmitry S. Koktysh[2,3], Daniel M. Fleetwood[1], Robert A. Weller[1], Robert A. Reed[1], and Sharon M. Weiss[*,1]

[1] Department of Electrical Engineering and Computer Science, Vanderbilt University, Nashville, TN 37212, USA;

[2] Vanderbilt Institute of Nanoscale Science and Engineering, Vanderbilt University, Nashville, TN 37212, USA

[3] Department of Chemistry, Vanderbilt University, Nashville, TN 37212, USA

Keywords: quantum dot; exciton; radiative lifetime; quenching; photoluminescence; defects; monolayer

ABSTRACT

Using a combination of continuous wave and time-resolved spectroscopy, we study the effects of interfacial conditions on the radiative lifetimes and photoluminescence intensities of colloidal CdTe/CdS quantum dots (QDs) embedded in a three-dimensional nanostructured silicon (NSi) matrix. The NSi matrix was thermally oxidized under different conditions to change the interfacial oxide thickness. QDs embedded in a NSi matrix with ~0.5 nm of interfacial oxide exhibited reduced photoluminescence intensity and nearly five times shorter radiative lifetimes (~16 ns) compared to QDs immobilized within completely oxidized, nanostructured silica ($NSiO_2$) frameworks (~78 ns). Optical absorption by the sub-nm oxidized NSi matrix partially lowers QD emission intensities while non-radiative carrier recombination and phonon assisted transitions influenced by defect sites within the oxide and NSi are believed to be the primary factors limiting the QD exciton lifetimes in the heterostructures.

a) Electronic mail: sharon.weiss@vanderbilt.edu

Colloidal quantum dots (QDs) possess a number of positive attributes, including a wide range of absorption and emission wavelengths, fast response times, and high rates of radiative emission. These attributes have made them attractive for use as energy emitters or absorbers in applications including LEDs[1-5], lasers[6,7], solar cells[8-14] and radiation scintillators.[15,16] For many of these applications, it is necessary for the QDs to maintain their optical properties when integrated with various materials and interfaces. Consequently, engineering non-radiative recombination rates to either suppress energy transfer and accelerate radiative emissions or enhance energy transfer to an interfacial material is of great interest.[17,18] Most studies to date investigate QD exciton interactions in multi-layer QD thin-films sandwiched between dielectric materials such as $TiO_2$ for energy conversion applications.[19,20] However, although there are reports discussing the role of surface defect states and ligands on QD recombination processes,[21] less attention has been given to understanding the influence of neighboring interfacial defects in nanostructured substrates. In this work, by achieving highly luminescent, monolayer QD distributions in large surface area, nanostructured silicon thin-films, we are able to straightforwardly probe the effects of interfacial defects located in close proximity to the QDs. This approach offers a significant advantage over most other techniques that utilize the deposition of several layers of QDs to achieve measurable signals, which can mask the influence of interfacial substrate defects in close proximity to the QDs. We first characterize the optical properties of monolayer CdTe/CdS QDs immobilized within nanostructured silicon/silicon dioxide (NSi/$SiO_2$) heterostructures and then vary the degree of interfacial oxide thickness to study the impact of the local environment on the optical properties of the immobilized QDs. The use of the high surface area (~200 $m^2$ $cm^{-3}$) NSi three-dimensional matrix as the embedding medium for monolayer QDs enables straightforward tuning of the interfacial composition from Si to $SiO_2$ through controlled thermal oxidation.

Completely oxidizing the NSi matrix to form a transparent nanostructured silica (NSiO$_2$) host matrix with a large conduction band offset leads to longer QD radiative lifetimes suitable for QD-integrated light emitting applications. In addition, use of a minimally oxidized NSi host matrix leads to enhanced exciton dissociation rates in the QDs, which may be favorable for photovoltaic applications if charged carriers can couple to lower energy sites present in the NSi.

NSi films were fabricated by electrochemical etching of boron doped p+ silicon wafers (<100>, 0.01Ω-cm, Silicon Quest) in a two-electrode configuration. The electrolyte consisted of an ethanolic HF solution (3:8 v/v 49-51% aqueous HF:ethanol, Sigma Aldrich). Anodization was carried out in the dark for 334 s at an etching current density of 48 mA/cm$^2$ to form 10 μm thick NSi films with average pore sizes of 25 nm. Each sample was rinsed thoroughly with ethanol and dried under a stream of nitrogen after the electrochemical etch. The samples were then cleaved in half to study the effects of thermal oxidation with each half subjected to either a 5 min, 500 °C oxidation or a 3 h, 1000 °C oxidation in air. In this way, the only difference between the samples was the oxidation parameters, which minimized any sample-to-sample variation that could result from NSi fabrication. The oxidized samples were incubated for 10 min in 3% polydiallyldimethylammonium chloride (PDDA) aqueous solution at pH = 3.0, followed by a deionized (DI) water rinse to remove excess molecules. PDDA molecules impart a positive charge to the oxidized NSi substrates upon attachment, which facilitates the monolayer assembly of colloidal QDs utilized for the studies conducted in this work. Visible light emitting CdTe/CdS QDs stabilized by glutathione were synthesized according to a procedure reported earlier.[22,23] The CdTe/CdS QDs were approximately 3.5 nm in diameter as estimated by transmission electron microscopy. Negatively charged CdTe/CdS QDs were electrostatically bound to the positively charged PDDA coated oxidized NSi surface during a 20 min incubation period to achieve monolayer QD distributions within the matrix. Unattached CdTe/CdS QDs were then washed away with thorough rinsing under DI water.

Absorbance and reflectance spectra were measured at room temperature with a Varian Cary 5000 UV-VIS-NIR spectrophotometer at a step size of 0.5 nm. Absorbance spectra were collected over a wavelength range of 300 nm - 800 nm. Reflectance spectra were collected over a wavelength range of 500

nm - 2000 nm using a spot size of ~6 mm. Continuous wave photoluminescence (CWPL) measurements were made using an Ar-Kr laser (Coherent Innova 70C) operating at a wavelength of 488 nm and power of 3 mW as the excitation source and a CCD spectrometer (Ocean Optics USB4000) fitted with a 1000 μm diameter optical fiber to record visible QD emission from the samples between 500 nm and 800 nm.

Based on Bruggeman effective medium theory and the measured blue-shift in the reflectance spectrum of the NSi film after the 500 °C oxidation step (not shown), we calculate the reduction in equivalent optical thickness to be 1191 nm and the resulting change in the effective refractive index of the NSi film to be ~0.06, which corresponds to approximately 0.05 nm of oxide growth. Fig. 1a shows the shifts in reflectance spectra for the NSi substrates following the attachment of PDDA molecules and QDs over a narrowed spectral range for clarity. Fig. 1b shows the absorbance and CWPL measurements of the QDs in solution. For the minimally oxidized NSi matrix, the observed red-shift of ~15 nm in the reflectance spectrum after QD immobilization corresponds to ~$10^{15}$ QDs captured within the NSi film and a near 8% surface area coverage, similar to what we reported in earlier work.[24] The smaller red-shift of ~9 nm that results from QD attachment to the $NSiO_2$ matrix can be explained by considering the additional oxide growth that reduces the effective pore size; the correspondingly reduced internal surface area in the $NSiO_2$ matrix leads to a reduction in the number of QDs attached [~$10^{14}$ QDs for a 1 – 2% surface area coverage]. In both the NSi and $NSiO_2$ matrices, we estimate inter-QD spacing to be greater than 10 nm, which is sufficient to suppress strong inter-QD carrier coupling.[25,26]

The measured CWPL spectra of the immobilized QDs in both substrates, as shown in Figs 1c-d, show good spectral agreement with the CWPL of QDs in solution (Fig. 1b). The intensities of the CWPL peaks scale with the number of QDs in the sample volume and, in the case of the QDs attached to the sub-nm oxidized NSi matrix, some of the QD emission is absorbed by the NSi. As reported previously, the spectral fringes readily seen in the CWPL spectrum in Fig. 1c, and not quite as visible in the spectrum in Fig. 1d due to the lower signal to noise ratio, confirm QD infiltration and immobilization throughout the embedding matrix and are a signature of Fabry-Perot interference.[24]

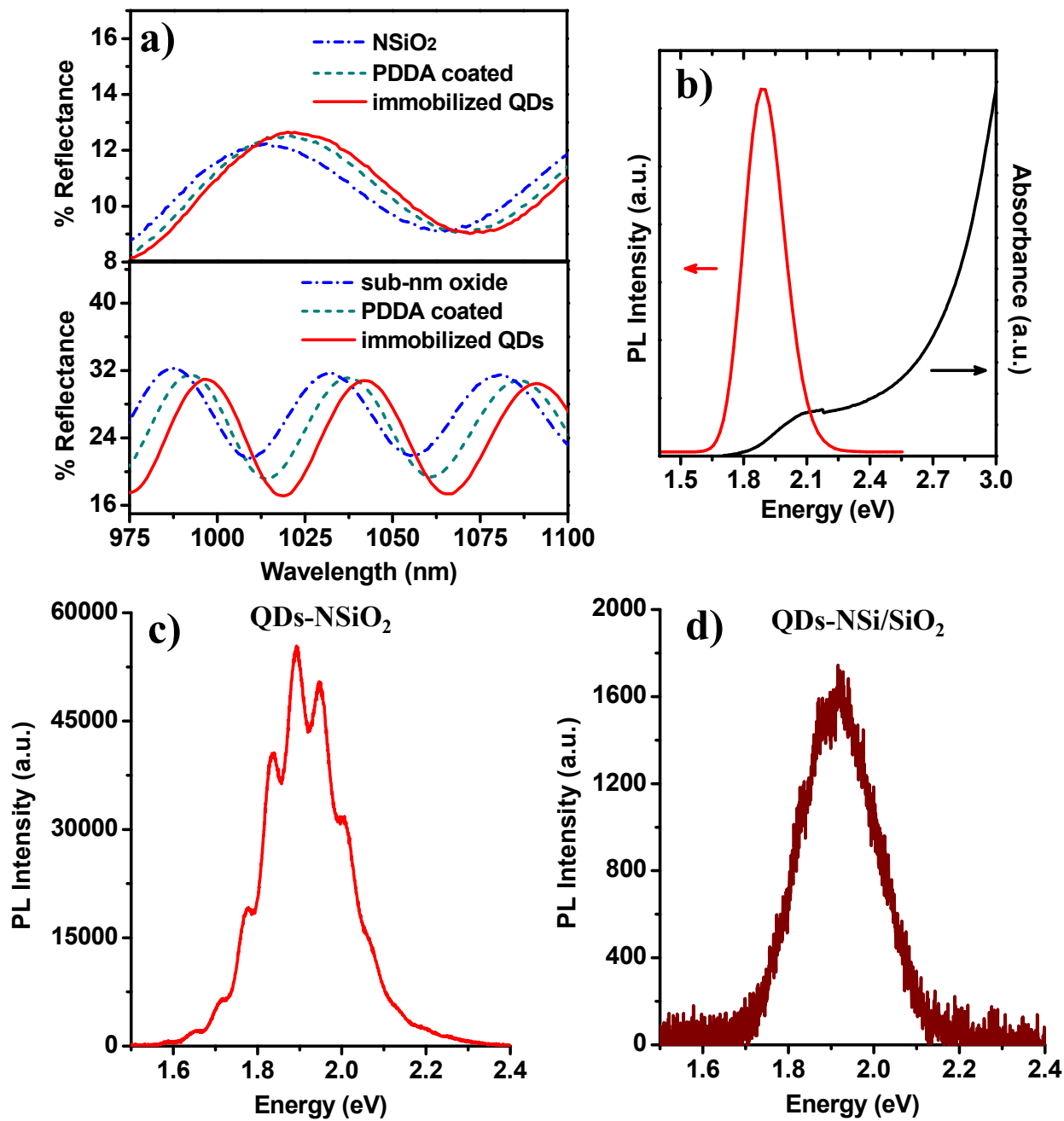

FIG. 1. a) Reflectance spectra measured for NSi thin-films following complete or partial oxidation, PDDA coating, and QD attachment. b) Absorbance and CWPL spectra for CdTe/CdS QDs in 0.1 mM aqueous solution. c) CWPL spectrum of monolayer QDs immobilized in a $NSiO_2$ thin-film. d) CWPL spectrum of monolayer QDs immobilized in sub-nm oxidized NSi thin-film.

Time-resolved photoluminescence (TRPL) measurements were carried out with an intensified CCD detector (iDUS490A, Andor Technology) attached to a spectrograph (Shamrock, SR303i, Andor Technology). A Nd:YAG Q-switched laser (Minilite-10, Continuum Inc.) operating at a wavelength of 355 nm in low power mode (10 mW) with 10 ns pulse duration and 10 Hz repetition rate was used as the excitation source for the TRPL experiments. Fig. 2 shows that QDs immobilized in NSiO$_2$ exhibit exciton lifetimes that are closely matched to those of QDs in solution, with decay times ($\tau$) of ~78 ns, consistent with prior work.[27] Hence, we conclude that the high-temperature grown oxide quality is largely defect free and excitons remain confined to the QDs themselves. Defect densities of the thermally grown SiO$_2$ have been extensively studied and, consistent with our conclusion, it has been shown that high temperature thermal oxidation results in a relatively low number of Si/SiO$_2$ interface trapped charges and oxide trapped charges with fewer sub-oxide species and primarily Si-SiO$_2$ groups (interface defect density <10$^{10}$ cm$^{-2}$ eV$^{-1}$).[28,29] Moreover, due to the unchanged exciton lifetime of QDs in solution and in the NSiO$_2$ matrix, the results in Fig. 2 suggest that the conformal PDDA coating in the oxidized NSi matrix does not influence QD exciton dynamics within the matrix and allows for uniform, monolayer attachment of the QDs.[30-32] The interfacial effects of the disordered nature of NSi, arising from varying nanocrystallite sizes, density of states, and defect levels at the band edges, become apparent when the QDs are immobilized in a sub-nm oxidized NSi framework, with exciton lifetimes decreasing by nearly five times to ~17 ns, as shown in Fig. 2. Prior studies suggest that low temperature oxide growth increases intrinsic stress and oxide densities.[33] Accordingly, the low temperature (500 °C) thermal oxidation of NSi likely results in a substantially higher interfacial defect density partially attributable to the presence of sub-oxides, Si-O-H groups with some Si-H bonds still present, intrinsic stress, and trapped charges. We previously reported that QDs immobilized in partially oxidized NSi substrates with ~2 nm of thermally grown oxide demonstrated exciton lifetimes close to ~34 ns, while a near ~3-4 nm of oxide results in exciton lifetimes closely matched to those of QDs in solution.[31] These results further support the correlation between QD exciton lifetime and the thickness and quality of NSi interfacial oxidation.

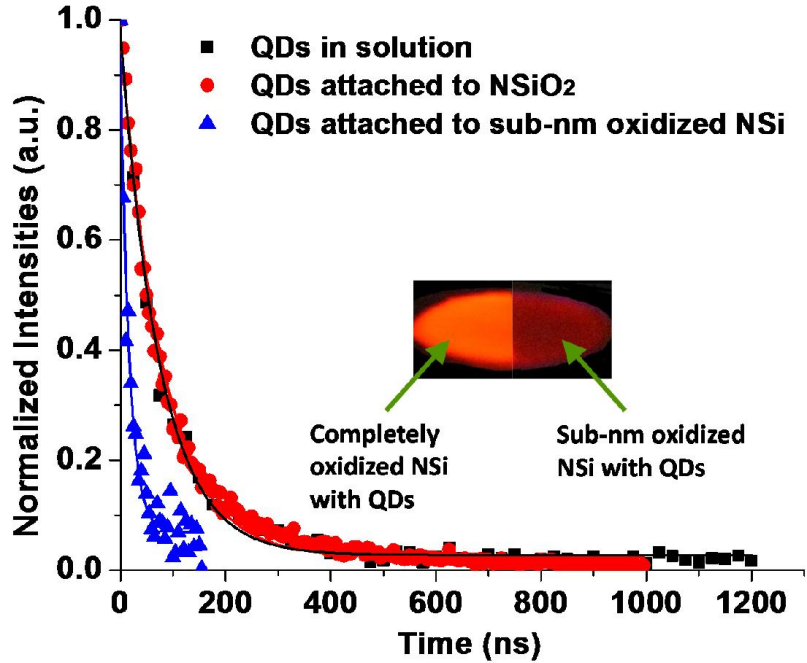

FIG. 2. Time-resolved photoluminescence measurements of QDs coupled with sub-nm oxidized NSi (solid triangle), QDs coupled with completely oxidized $NSiO_2$ (filled circles), and QDs in solution (filled squares) with solid lines indicative of a single exponential decay. The inset camera image shows a NSi substrate that has been cleaved in half, with each half subjected to a different oxidation condition: one half is oxidized at 500 °C for 5 min, forming a sub-nm oxide, and the other half is completely oxidized after oxidation at 1000 °C for 3 h.

To provide insight into the mechanisms responsible for the different QD exciton lifetimes in NSi matrices with different interfacial oxide quality and thickness, we propose possible exciton quenching pathways, which are shown in Fig. 3 along with the relative energy levels for the QDs and embedding matrices. The energy levels are based on those reported in the literature for NSi and CdTe/CdS QDs.[34-39] Prior studies have suggested that CdTe/CdS QDs may exhibit either Type I or Type II behavior depending upon core and shell thickness and the influence of lattice strain.[40] Discrepancies in the band offset calculations between CdTe and CdS can also change the suggested classification of the heterostructure.[27,34] The type of QD emission observed in our experiments is mostly Type II centered near 1.85 eV. Based on the relative position of the energy levels, the holes are expected to be strongly confined within the core while the electrons are most likely delocalized over the CdS shell and, in the case of the sub-nm oxidized

NSi substrate (Fig. 3b), may couple into the NSi region via an energetically favorable transition. As a result, exciton dynamics are greatly affected by electron wave function overlap with surface and substrate states. Since the inter-QD separations estimated in the NSi frameworks suggest minimal QD-QD exciton coupling will occur, interfacial effects between the QDs and embedding matrix are expected to dominate the QD exciton dynamics.[41,42] The reduced exciton lifetime for the QDs immobilized within the NSi matrix with low quality, ultra-thin oxide is probably due to exciton quenching, either through non-radiative carrier recombination or phonon assisted transitions influenced by defect sites within the oxide as well as the high density of disordered states present in the NSi substrates where defect densities may be close to $10^{16}$ cm$^{-2}$ (Fig. 3b).[43,44] We note that many studies of the non-radiative phonon transition processes related to NSi have been performed.[38, 45] If the QD-NSi system can be engineered to promote carrier migration from the excited states of the QDs to the underlying NSi substrate with minimal trapping in oxide defects, perhaps through the use of a higher quality thin oxide grown by atomic layer deposition, then the system could be potentially useful for energy storage and solar cell applications. Photocurrent studies investigating the role of a thin high quality oxide at the interface would be able to provide greater insights into charge transport within such complex structures. For the case of QDs attached to completely oxidized $NSiO_2$ substrates, the prolonged high temperature oxidation is expected to result in lower carrier trap sites within the immediate vicinity of the QDs. The thicker oxide also prohibits tunneling of carriers to NSi disordered states. As a result, the dominating photo-physical mechanisms for QDs in $NSiO_2$ are exciton generation and radiative recombination (Fig. 3a). Therefore, $NSiO_2$ based optical structures such as microcavities and distributed Bragg reflectors with in-built QD emitters are a potentially advantageous platform for QD based light emitting applications.[42] While the work presented here provides insights into the importance of the interface in colloidal QD-integrated heterostructure devices, further experiments that can resolve exciton lifetimes below 10 ns and therefore allow for the possibility of multi-exponential fits are needed to more definitively elucidate the radiative and non-radiative mechanisms involved in QD exciton decay within three-dimensional heterostructures.

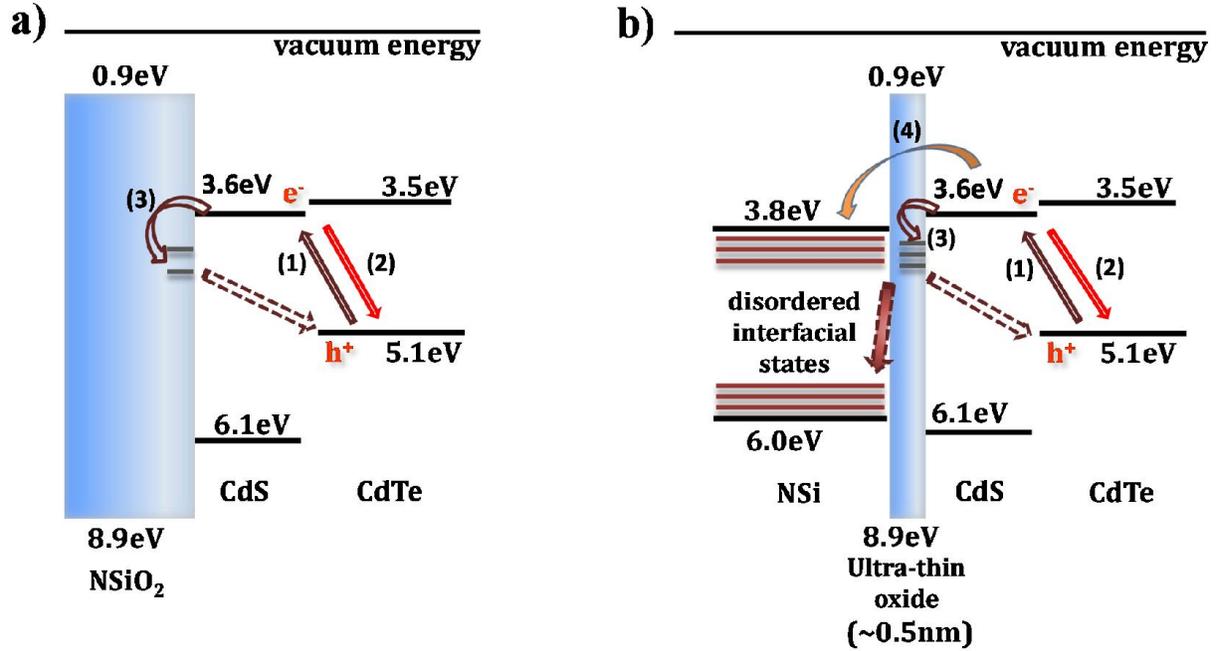

FIG. 3. Energy level representations of CdTe/CdS QDs immobilized in a) completely oxidized $NSiO_2$ and b) NSi with ultra-thin oxide. Possible photo-physical mechanisms are: (1) absorption of photon energy and creation of an exciton. (2) Radiative decay of exciton through CdTe/CdS band-edge emission. (3) Migration of an exciton to lower energy sites in the interfacial oxide, causing exciton dissociation. The mobile electron may then remain trapped for extended periods of time in oxide defect states or non-radiatively decay either to the CdTe/CdS valence band or relax to the NSi valence band in b) via coupling to the disordered NSi interfacial states and decaying via phonon modes. (4) The energetically favorable transition of electrons from the CdTe core to the NSi interfacial states in b). The energy levels shown are based on those reported in [34-39].

In summary, we demonstrate the importance of local environments and interfacial materials on QD fluorescence emission and exciton dynamics. By altering the coupled QD material interface from a sub-nm oxidized NSi to a completely oxidized $NSiO_2$ framework, we are able to significantly suppress non-radiative recombination pathways of photogenerated excitons and achieve nearly 5 times longer exciton lifetimes of ~78 ns that are on par with QDs in solution. This work contributes to the growing field of QD-based excitonic devices that are gaining increasing attention for applications in photovoltaics, solar cells, and light sources. Further theoretical calculations of exciton localization and transfer mechanisms in the NSi-QD heterostructures and temperature dependent, time-resolved spectroscopic studies could help provide a greater understanding of charge transfer and exciton confinement within QD integrated devices.

This work was funded in part by the Defense Threat Reduction Agency (Grant no. HDTRA1-10-1-0041). The authors thank the Vanderbilt Institute for Nanoscale Science and Engineering for equipment use and facilities that were renovated under NSF ARI-R2 DMR-0963361. The authors gratefully acknowledge Prof. J. E. Macdonald for providing insightful suggestions, S. Avanesyan for assistance with the pulsed laser setup, and S. M. Harrell for useful technical discussions.